\documentclass[9pt,twocolumn,twoside]{osajnl}

\journal{ol} 

\newcommand{\ie}{{\em i.e.}}

\setboolean{shortarticle}{true}

\title{A realistic phase screen model for forward multiple-scattering media}

\author[1]{Mu Qiao}
\author[2,*]{Xin Yuan}

\affil[1]{Department of Electrical and Computer Engineering, New Jersey Institute of Technology, Newark, New Jersey, 07102, USA}
\affil[2]{Nokia Bell Labs, 600 Mountain Avenue, Murray Hill, NJ, 07974, USA}
\affil[*]{Corresponding author: xyuan@bell-labs.com}

\begin{abstract}
	{Existing random phase screen (RPS) models for forward multiple-scattering media fail to incorporate ballistic light. 
	In this letter, we redesign the angular spectrum of the screen by means of Monte-Carlo simulation based on an assumption that a single screen should represent all the scattering events a photon experiences between two adjacent screens.
	Three examples demonstrate that the proposed model exhibits more realistic optical properties than conventional RPS models in terms of attenuation of ballistic light, evolution of beam profile and angular memory effect.
	The proposed model also provides the flexibility to balance the computing accuracy, speed and memory usage by tuning the screen spacing. \textbf{}}
\end{abstract}

\setboolean{displaycopyright}{true}

\begin{document}
	
	\maketitle
	\begin{figure*}[!htbp]
		\centering
		\fbox{\includegraphics[width=0.95\linewidth]{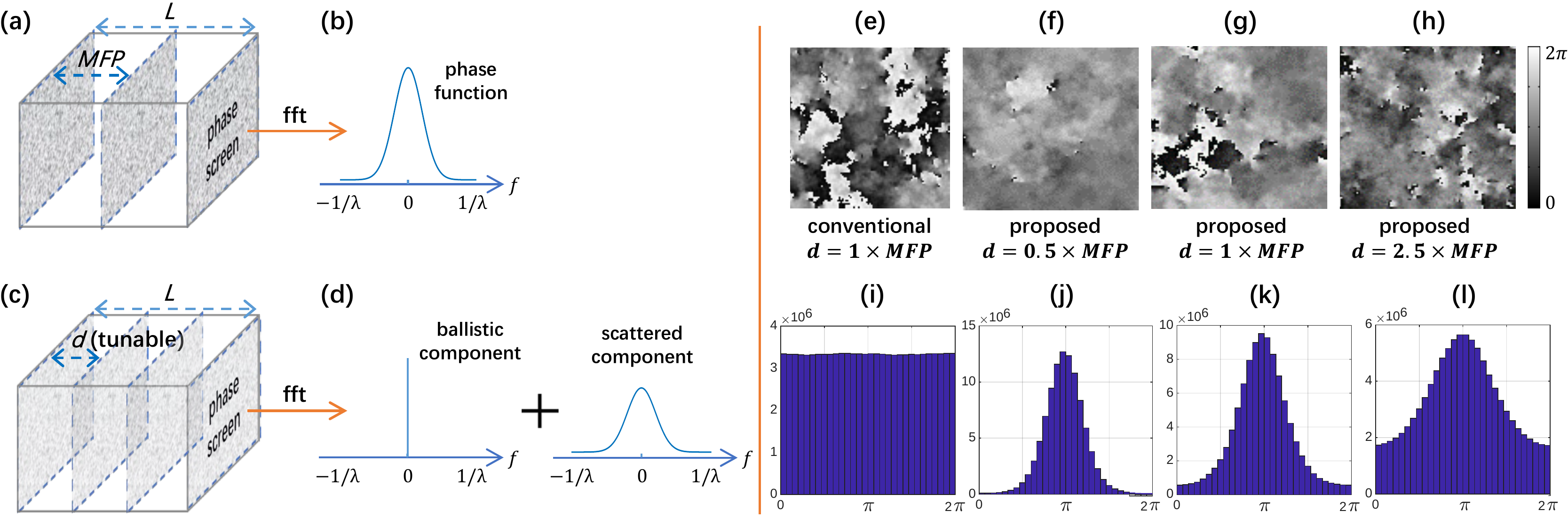}}
		\caption{Comparison between the principles of our proposed model (left part, bottom row) and the conventional RPS model (left part, top row). In the conventional model, the screen spacing $d$ must be the MFP of the medium to be modeled (a). The Fourier transform (b) (plotted along either $f_x$ or $f_y$ axis) of the screen is determined by the phase function of a single scattering event. In the proposed model (c), $d$ can be set arbitrarily. The Fourier transform of the screen is determined by Monte-Carlo method, and consists of two parts: the zero-order part (d, left) which representing the ballistic light and the high-order part (d, right) representing the scattered light. 
		(e-h) Screen samples of $100\times100$ pixels (each pixel represents a $0.2{\mu}m\times0.2{\mu}m$ area) in the conventional model ($d=1\times$MFP) and the proposed model when $d$ is set to $0.5\times$MFP, $1\times$MFP and $2.5\times$MFP, respectively. 
		(i-l) Phase histograms of (e-h) ($10000\times10000$ pixels considered)
		}
		\label{fig:Model}
	\end{figure*}
	Random Phase Screen (RPS) models~\cite{goodman2015statistical,schott2015characterization} are designed to simulate coherent wave propagation through scattering media, and thus have been a powerful tool in the analysis and simulation of emerging interference-related optical imaging techniques, such as optical phase conjugation~\cite{yaqoob2008optical}, transmission matrix~\cite{popoff2010measuring}, wavefront shaping~\cite{horstmeyer2015guidestar}, and those techniques based on speckle correlation~\cite{feng1988correlations,freund1988memory,bertolotti2012non,qiao2018bidirectional,qiao2017non,osnabrugge2017generalized}.
	However, the existing RPS model~\cite{schott2015characterization} (to the best of our knowledge) fails to incorporate the ballistic light because its screen modulation function only includes a negligible fraction of zero-frequency component. Thereby, nearly all the incident light will be scattered even after a single screen, which is unrealistic in practical applications. 
	In order to fill this gap, we propose a new method in this letter to design the screen modulation function.
	
	To be concrete, we start from the principle of the RPS model.
	In the RPS framework, 
	a scattering medium is modeled by a number of equally-spaced random phase screens (Fig.~\ref{fig:Model}(a, c)) and scattering is considered to occur only on these discrete screens; light propagation between screens is modeled by free-space diffraction. Therefore, the process of light transmitting through the media is alternated between free space propagation and screen modulations. The free space propagation can be calculated by the Rayleigh-Sommerfeld diffraction method or the angular spectrum diffraction method. The modulation of the screen on the incident light field can be expressed as:
	\begin{equation}
	E_k^+(x,y)=E_k^-(x,y) M_k(x,y),  \label{Eq:EoEi}
	\end{equation}
	where $(x,y)$ represents spatial coordinate, $E_k^-(x,y)$ and $E_k^+(x,y)$ represent the light field immediately before and after the $k$th screen, respectively, and $M_k(x,y)=e^{i\Phi_k(x,y)}$ signifies the complex transmission function of the $k$th screen with $\Phi_k(x,y)$ denoting its phase distribution. Applying Fourier decomposition to $M_k(x,y)$, we have
	\begin{eqnarray}
	E_k^+(x,y)&=& E_k^-(x,y)	\int_{-1/\lambda}^{1/\lambda}\int_{-1/\lambda}^{1/\lambda}{\cal F}_k(f_{x},f_{y}) \nonumber\\
	&& \qquad \qquad~~ \times e^{i2\pi(f_{x} x+f_{y} y)}df_{x}df_{y},
	\label{Eq:E0_F}  
	\end{eqnarray}
	where $\lambda$ denotes the wavelength of the light, $(f_{x},f_{y})$ represents spatial frequency, and ${\cal F}_k(f_{x}, f_{y})$ is the Fourier transform of $M_k(x,y)$.
	Each Fourier component $e^{i2\pi(f_{x} x+f_{y} y)}$ deflects the input field $E_k^-$ towards different polar angle $\theta$ and azimuth angle $\phi$, where
	\begin{equation}
	\theta=\arcsin (\lambda \sqrt{f^2_{x} + f^2_{y}} ), \quad \phi = \arctan(f_{y}/f_{x}).
	\label{Eq:3}
	\end{equation}
	Converting ${\cal F}_k(f_{x},f_{y} )$ from the {$(f_{x},f_{y})$} coordinate to the $(\theta,\phi)$ coordinate, we obtain the complex scattering angle distribution $S_k(\theta,\phi)$ of the screen, which represents the complex amplitude of the scattered light field when a plane wave is incident.
	Note that the Fourier components with frequencies beyond the integral interval $(-1/\lambda,1/\lambda)$ in Eq.~(\ref{Eq:E0_F}) represent fast-decaying evanescent waves, which are ignored here because the thickness of the free space between adjacent screens (sub-millimeter scale) is much longer than the decay length of the evanescent wave (several wavelengths).
	
	The RPS model has three independent parameters: screen spacing $d$, number of screens $N$, and the aforementioned complex scattering angle distribution $S(\theta,\phi)$.
	To map these parameters to the properties of real scattering media which consist of a huge number of scattering particles, the conventional method~\cite{schott2015characterization} first assumes that each screen represents {\em a single scattering event}, which refers to the event that a photon runs into a scattering particle. 
    Then, the intensity distribution of the scattered light $|S(\theta,\phi)|^2$ after a single screen should be the same as that of the photon after a single scattering event.
	This angle distribution is termed {\em phase function} $P(\theta)$ in Mie theory, and is related to the anisotropy factor {\em $g$} of the medium by~\cite{wang2012biomedical}:
	\begin{equation}
	P(\theta) = \frac{1-g^2}{2(1+g^2-2g\cos\theta)^{3/2}}. \label{Eq:Ptheta}
	\end{equation}
	The screen spacing $d$ (Fig.~\ref{fig:Model}(a)) is then necessarily set to the scattering {\em mean free path} (MFP) of the medium, which measures the average photon path length between two consecutive {\em single scattering events}.

	Though it seems reasonable to assume that a single screen represents a single scattering event and correspondingly set the screen function as the phase function, there is a fundamental drawback in this model: {\em the ballistic light (unscattered light component) is not effectively incorporated into the model}. 
	
	From~\eqref{Eq:E0_F} we know that the zero-frequency component of the screen function dose not impose any modulation to the incident light field $E_k^-$. 
	Therefore, the proportion of the zero-frequency component in the screen function represents the proportion of ballistic light in the transmitted field  $E_k^+$. 
	However, since the phase function $P(\theta)$ in~\eqref{Eq:Ptheta} is continuous, it only has a negligible fraction at the exact zero-frequency position (Fig.~\ref{fig:Model}(b)). In this case, nearly all the incident light will be scattered even after a single screen. However, for real scattering media, the attenuation of the ballistic light follows the Beer's law:
	\begin{equation}
	I_b = I_i e^{-z/\ell_s}, \label{Eq:Beer}
	\end{equation}
	where $I_i$ represents the intensity of the incident light and $I_b$ signifies the intensity of the ballistic light after propagating a distance $z$ in a medium with an MFP of $ell_s$. According to \eqref{Eq:Beer}, after one MFP (the distance between adjacent screens in the above conventional RPS model), there is still an $e^{-1}$ fraction of ballistic light left, which is a significant amount. 
	
	\begin{figure}[!b]
		\vspace{-4mm}
		\centering
		\fbox{\includegraphics[width=.6\linewidth]{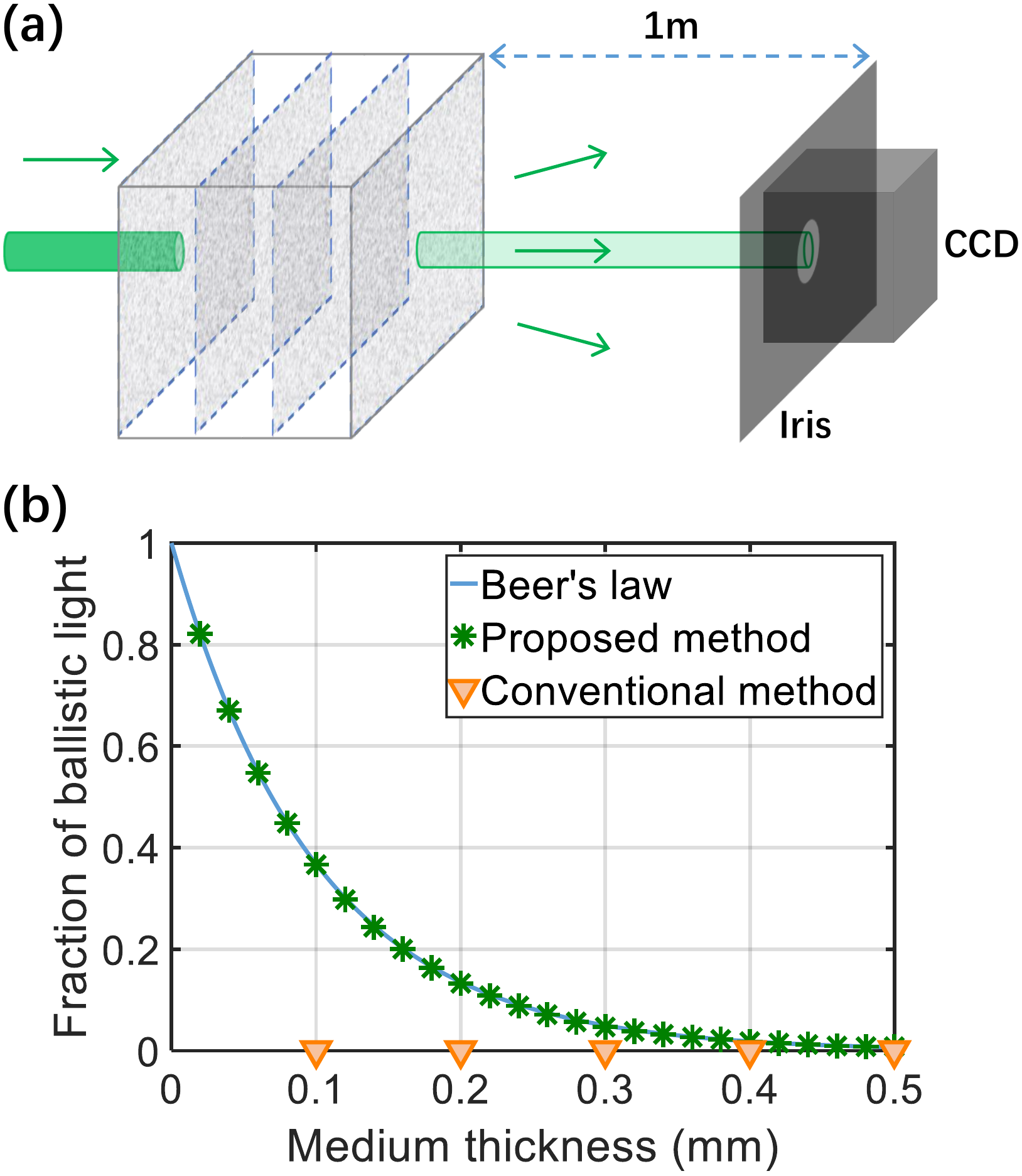}}
		\caption{Attenuation of ballistic light with increasing medium thickness. (a) Illustration of the method of measuring the intensity of ballistic light in our numerical experiments. The incident Gaussian beam has an FWHM diameter of $0.35mm$ and the sample has a $0.1mm$ MFP. To reduce the influence of the scattered light, the CCD (recording plane) is placed far ($1m$) away from the medium to dissipate the scattered light sufficiently; then an iris is inserted right before the CCD to select the ballistic light. (b) Results obtained using the proposed model (green-star markers) and the conventional model (orange-triangle markers). The blue curve is the theoretical prediction by the Beer's law.}
		\label{fig:Attenuation}
	\end{figure}

	\begin{figure*}[!t]
		\centering
		\fbox{\includegraphics[width=.75\linewidth,height = 7cm]{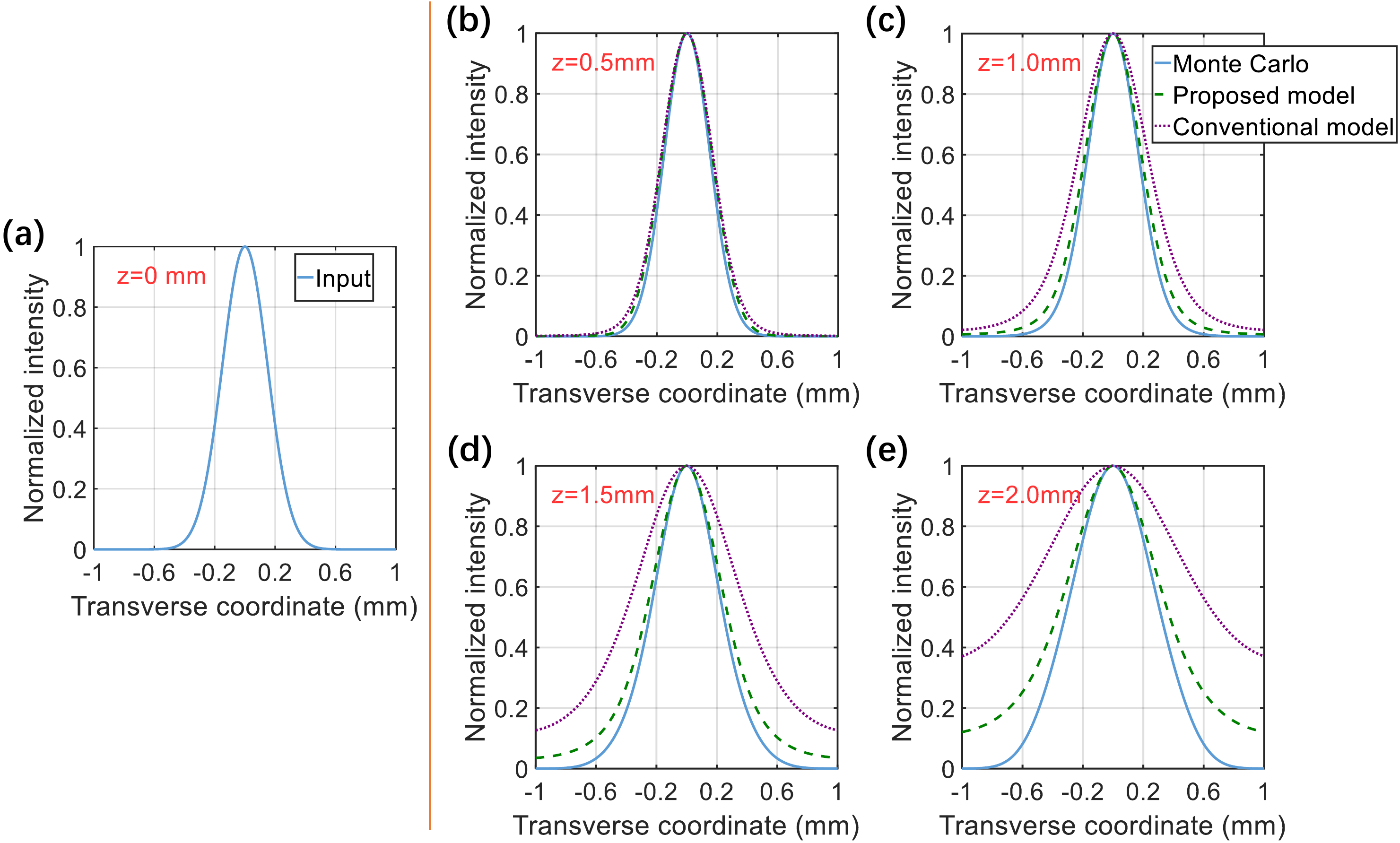}}
		\vspace{-3mm}
		\caption{Expansion of a Gaussian beam inside a scattering medium of an MFP of $0.1$ mm and an anisotropy factor of $0.98$. (a) Transverse intensity profile of the input Gaussian beam which has an FWHM diameter of $0.35$ mm. (b-e) Beam profiles at different depths $z$. The blue-solid, green-dash and purple-dot lines are obtained using the Monte Carlo method (which is the truth), our proposed and the conventional RPS model, respectively. The curves are smoothed by average over the azimuth angle $\phi$ since both the input Gaussian beam and the scattering angle distribution are circularly symmetric.}
		\vspace{-3mm}
		\label{fig:Expansion}
	\end{figure*}

	In order to fill this gap, in this letter, we propose a new method to design the screen function, which takes into account the contributions of both the ballistic and the scattered light to the screen function. 
	Our method is based on the following assumption: if the {screen spacing} is set to $d$, which can be arbitrary, then a single screen must represent the summation of the events a photon experienced after propagating the distance $d$ in the medium to be modeled. 
	Therefore, the intensity distribution of the scattered light $|S(\theta,\phi)|^2$ after a single screen  should be the same as the overall angle distribution of the photon after these events, which include ballistic propagation, single scattering and multiple scattering. 
	This distribution can be obtained either by Mie theory or Monte Carlo simulation~\cite{Wang95MCMC}.  
	Here, we choose the later one since the Monte Carlo model is simple and effective.

	In our Monte Carlo simulation to determine the scattering angle distribution of the photon, we first send a pencil beam (which is an ideal beam with an infinitesimal beam size) containing 10 billion photons incident normally on the medium, and then record the distribution of their directions after they arrive at depth $d$. Since the phase function in a single scattering event is circularly symmetric (depends only on the polar angle $\theta$), the direction distribution of the scattered photons is also circularly symmetric. Thereby, we conducted an average along the azimuth angle $\phi$ to obtain a smooth statistical direction distribution of these photons. The obtained distribution $|S(\theta,\phi)|^2$ consists of two parts: the zero-frequency part representing ballistic light (Fig.~\ref{fig:Model}(d), left), and the high-frequency part representing scattered light (Fig.~\ref{fig:Model}(d), right).

	After obtaining the scattering angle distribution $|S(\theta,\phi)|^2$ of the screen, by performing a coordinate transform from the $(\theta,\phi)$ to the $(f_x ,f_y)$ coordinate according to ~\eqref{Eq:3}, we obtain the amplitude of the Fourier transform of the screen $|{\cal F}_k(f_{x},f_{y})|$.
	To completely define the screen modulation function, we also need to know the phase distribution of ${\cal F}_k$. This can be obtained by the Gerchberg-Saxton algorithm~\cite{fienup1982phase} using the constrain that $M(x,y)$ is a phase only function (which means it has a uniform amplitude distribution).

    In the above analysis, we only considered the statistical property of the screen, \ie, the smooth profile of the scattering angle distribution $|S(\theta,\phi)|^2$ (See Fig. 1d, except the zero point). However, in nature,  scatterings happen in an irregular manner. To embody this irregularity in our model,  we superpose a random modulation $R(\theta,\phi)$ (with value ranging from 0 to 1) onto the statistical distribution $|S(\theta,\phi)|^2$ before implementing the above procedure to calculate the $M(x,y)$. Each individual screen in the model should be calculated using different $R(\theta,\phi)$ to avoid spatial periodicity.
	
	Our redesigned screen modulation function endues the RPS model with more realistic optical properties than the conventional method described above. In the following, we demonstrate this superiority through three examples.

	Since the most distinctive feature of our proposed model is the integration of the ballistic light, we first investigate the attenuation of the ballistic light when the thickness of the scattering medium is gradually increased. The medium in our simulation has an MFP of $0.1$ mm and an anisotropy of $0.98$ for the employed $532$ nm light. We send a Gaussian beam with a full width at half maximum (FWHM) diameter of $0.35$ mm into the medium and then record the intensity of the transmitted (ballistic) Gaussian beam on the other side of the medium (Fig.~\ref{fig:Attenuation}(a)). Since the transmitted Gaussian beam is mixed with scattered light, we place the recording plane far ($1$ m) away from the medium to dissipate the scattered light so that its intensity within the range of the Gaussian beam is negligible compared to that of the Gaussian beam. We then place an iris concentric with the Gaussian beam to block the scattered light and only record the intensity within the iris. Fig.~\ref{fig:Attenuation}(b) shows the results obtained from our model (green-star marks, screen spacing $d$ is set as $0.2\times$MFP) and the conventional model (orange-triangles marks). The prediction from the Beer’s law is also plotted (blue-solid curve) as the truth. It can be observed that our model agrees well with the Beer’s law. By contrast, in the conventional model, no ballistic light will be left after one MFP.

	Next, we provide two examples to show that our model exhibits closer optical properties to real scattering media. First, we investigate the expansion of a Gaussian beam inside a medium. The medium and the incident Gaussian beam have the same parameters as in the above test. In this example, we use the results from the Monte Carlo method as the truth. Fig.~\ref{fig:Expansion}(a) shows the one-dimensional profile of the incident Gaussian beam. Fig.~\ref{fig:Expansion}(b)-(e) show the beam profiles at different depths, which are smoothed by average over the azimuth angle $\phi$. It can be observed that as the depth increases, our model follows the truth well, but the conventional model exhibits much broader profiles especially at large depths. The slightly larger tails in our model than those in the MC model is caused by the following two factors. First, in the RPS simulation, light field is represented by a 2D matrix with a finite size. This imposes a constrain on the area of the light field and therefore causes diffraction of the light field at the edges, hence the tails. A larger matrix size can reduce this effect. Secondly, in the MC simulation, the incident Gaussian beam is modeled by a bundle of pencil beams incident normally onto the media, which means the diverging angle of the incident beam is zero. However, in the RPS model, the incident Gaussian field has a finite size and a finite bandwidth of its spatial frequency, which means a larger diverging angle and thus faster beam expansion.
	
\begin{figure}[!tb]
		\centering
		\fbox{\includegraphics[width=.95\linewidth]{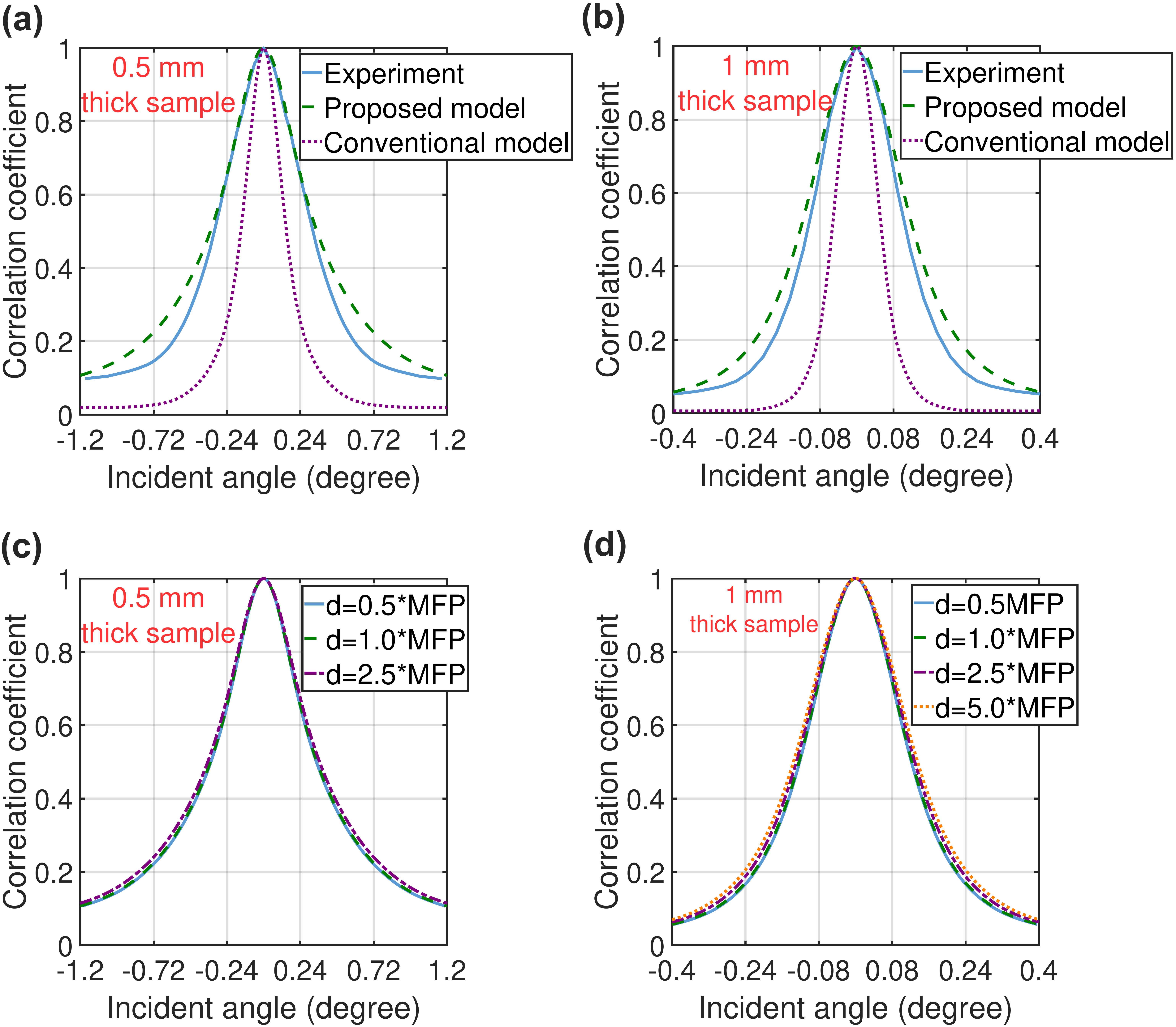}}
		\vspace{-3mm}
		\caption{Angular memory effect curves. (a,b) experimental (blue-solid curve) and simulation results obtained using the proposed (green-dash curve) and the conventional (purple-dot curve) phase screen models, for $0.5$mm (a) and $1.0$mm (b) thick samples with MFP$=0.1mm$ and $g=0.98$. The $x$-axis is the incident angle of the Gaussian beam and the $y$-axis is the correlation coefficient between the initial and the tilted output speckle field. (c,d) memory effect curves obtained with different setting of $d$ in the proposed model.}
		\vspace{-3mm}
		\label{fig:memory}
	\end{figure}

	Now, we investigate the angular memory effect (AME)~\cite{feng1988correlations,freund1988memory}, which has been widely employed in novel imaging techniques~\cite{bertolotti2012non,qiao2018bidirectional,qiao2017non}. In this test, we use experimental results as the truth. In the experiment, we create the scattering medium by dispersing silica microspheres (scattering particles) into transparent gelatin solutions. The microsphere has a $4.3 \mu$m diameter, a $1.45$ refractive index and a $2\times10^{-4}/\mu m^3$ concentration, and the gelatin solution has a $1.33$ refractive index. These parameters yield an MFP of $0.1mm$ and an anisotropy factor $g$ of $0.98$ for $830nm$ light according to the Mie theory. We measure the AME curve by tilting the angle of incidence of a Gaussian beam with $0.6mm$ diameter and then calculate the correlation between the initial output speckle field (obtained at zero degree incidence) and the tilted output field at each incident angle.
	The results are shown in Fig.~\ref{fig:memory}(a,b) by the blue-solid curves. 
	We then numerically repeat the above experiments.
	The results obtained with our and the conventional model are shown in Fig.~\ref{fig:memory}(a,b) by the green-dash and purple-dot curves, respectively. 
	For both $0.5$ mm (Fig.~\ref{fig:memory}(a)) and $1$ mm (Fig.~\ref{fig:memory}(b)) thick media, our model is closer to the experimental results both in terms of the width of the correlation curve and the asymptote at large tilting angles, which arises from the correlation of the ballistic light. The slightly larger AME range of our model than the experimental results might arise from the imperfect experimental conditions, such as decorrelation of the medium (drifting of the scattering particles inside the medium) during the measurement. Fig.~\ref{fig:memory}(c) and Fig.~\ref{fig:memory}(d) compares the angular correlation curves obtained with different mask spacing $d$ in our model, from which we observe that the model is largely independent of the choice of $d$. However, to correctly model the memory effect, at least two masks~\cite{haskel2018modeling} (with $d=0.5*thickness$) are required since a single mask (with $d=thickness$) will exhibit an infinitely large correlation range, which is obviously unrealistic.
	
	We foresee two kinds of potential applications of our proposed model, which cannot be implemented by the conventional RPS model. 
	First, the incorporation of ballistic light makes our model capable to evaluate the performance of ballistic imaging techniques ~\cite{kang2015imaging}.
	Second, the screen spacing $d$ in our model can be set flexibly to meet different requirements in different applications. For example, it can be set smaller than MFP to resolve scattering process within one MFP, or bigger than MFP to reduce the calculation time and memory usage when the lost accuracy is acceptable. 
	Furthermore, since the medium thickness in the RPS model can only be integral multiples of the screen spacing $d$, our model can handle the thickness more precisely by choosing a smaller $d$. The common drawback of the RPS model still remains: back-scattered light is not considered, though recent efforts~\cite{tahir2019holographic} have attempted to address this issue.
	
	In summary, we have proposed a new method to design the screen function in the RPS model, which incorporates the ballistic light into the model, endues the model with more realistic optical properties, and provides the flexibility to balance computing resources and precision.
	
\noindent{\bf Acknowledgments.} M. Qiao acknowledges the funding support from Nokia Bell Labs to build the hardware setup.

\noindent {\bf Disclosures.} MQ: Nokia (F), XY: Nokia (E).

\end{document}